\documentclass[12pt]{article}

\usepackage{amsmath}
\def\be{\begin{equation}}
\def\ee{\end{equation}}
\begin{document}

\begin{center}
{\Large \bf A fifth order differential equation for charged perfect fluids}\\
\vspace{1,5cm}{\bf M. C. Kweyama $^\dag$, K. S. Govinder  and S. D. Maharaj $^\ddag$} \\
Astrophysics and Cosmology Research Unit,\\ School of Mathematical Sciences,\\ University of KwaZulu-Natal,\\
 Private Bag X54001,\\ Durban 4000,\\ South Africa.\\
\vspace{1,5cm}{\bf Abstract}\\
\end{center}
We investigate the master nonlinear partial differential equation that governs the evolution of shear-free spherically symmetric charged 
fluids.  We use an approach which has not been considered previously for the underlying equation in shear-free spherically symmetric 
spacetimes.  We derive a fifth order purely differential equation that must be satisfied for the underlying equation to admit a Lie point 
symmetry.  We then perform a comprehensive analysis of this equation utilising the Lie symmetry analysis and direct integration.  This 
enables us to reduce the fifth order equation to quadratures. Earlier results are shown to be contained in our general treatment.\\
 ~\\
 Keywords: Einstein-Maxwell equations; exact solutions; charged relativistic fluids\\
~\\
PACS numbers: 02.30.Jr, 04.20.Jb, 04.40.Nr\\
 ~\\
$^\dag$Permanent address: Department of Mathematical Sciences, Mangosuthu University of Technology, P. O. Box 12363, Jacobs, 4026, 
Durban, South Africa
 ~\\
$^\ddag$ Correspondence to

\newpage

\section{Introduction}

The Einstein-Maxwell system of equations plays a crucial role in relativistic astrophysics when describing
spherically symmetric gravitational fields in static manifolds.  In these situations we are modeling  charged
 compact objects with strong gravitational fields such as dense neutron stars.  Several recent treatments, including the works of
Ivanov \cite{Ivanov} and Sharma \textit{et al} \cite{Sharma},
demonstrate that the presence of the electromagnetic field affects
the values of redshifts, luminosities and maximum mass of a compact
relativistic star.  The presence of electric charge is a necessary ingredient in the structure and gravitational evolution
  of stars composed of quark matter as illustrated in the treatments of Mak and Harko \cite{Mak} and Komathiraj and
  Maharaj \cite{Komathiraj}.  Therefore exact models describing the formation  and evolution of charged stellar objects, within
 the context of full general relativity, are necessary.  Lasky and Lun \cite{Lasky, Lun}, amongst
   others, studied the role of electromagnetic fields in gravitational collapse, formation of black holes and the existence
   of naked singularities.  Significant electric fields are also present
 in phases of intense dynamical activity, in collapsing configurations, with time scales of the
 order of the hydrostatic time scale for which the usual stable equilibrium configuration assumptions
  are not reliable (as shown in the treatments of Di Prisco \textit{et al} \cite{Di} and Herrera \textit{et al} \cite{Herrera}).

When seeking exact solutions to the Einstein field equations for an uncharged case, the common practice is to make some simplifying 
assumptions.  The assumptions involve shear-free and spherically symmetric spacetimes.  This approach allows for the reduction of the field 
equations to a single second order partial differential equation.  It is remarkable that the resulting equation can be treated as an ordinary 
differential equation.  For this case the first general class of solutions obtained was a result of the treatment by Kustaanheimo and Qvist \cite
{Kustaanheimo}.    Srivastava \cite{Srivastava} and Sussman \cite{Suss89} provide extensive treatments of the uncharged case.  The 
generalisation to include the electromagnetic field is easily performed and is described by the Einstein-Maxwell system of equations. Again in 
this case the field equations can be reduced to a single second order differential equation which, interestingly, can also be treated as an 
ordinary differential equation.  Krasinski \cite{Krasinski} presents a review of known solutions for the charged case that admit a Friedman 
limit.  A comprehensive study of the physical and mathematical properties of the Einstein-Maxwell system in spherical symmetry has been 
performed by Sussman \cite{Suss88a, Suss88b} and Srivastava \cite{SrivaDC}.

The main objective of this paper is to study the integrability features of the underlying differential equation for the Einstein-Maxwell system of 
field equations.  In section 2, we reduce the system of Einstein-Maxwell field equations, generalising the transformation of Faulkes \cite
{Faulkes}, to a single nonlinear second order partial differential equation.  This is a governing equation modeling the evolution of shear-free 
spherically symmetric charged fluids.  The conventional approach is to use different
mathematical techniques to solve the master second order equation. 
The recent treatment of Kweyama \textit{et al} \cite{KMG} is a comprehensive attempt
to investigate the conditions for the equation to admit a first integral or be reduced to
quadratures using the Lie symmetry methods of differential equations.
Here we consider a different approach which we believe is unique and has not been considered before for 
the governing equation in shear-free spherically symmetric spacetimes.  For the existence of a Lie point symmetry we derive, in section 3, a 
fifth order purely differential equation that must be satisfied.  A detailed analysis of this equation is performed.  We solve, for the first time, 
the relevant integro-differential equation arising from the integration of the fifth order differential equation.  A brief discussion of the results is 
given in section 4.

\section{Field equations}

We consider the shear-free motion of a spherically symmetric perfect
fluid in the presence of the electromagnetic field.  We choose a
coordinate system $x^{i}=(t,r,\theta,\phi)$ which is both comoving
and isotropic.  In this coordinate system the metric can be written
as
\begin{equation}
ds^{2}=-e^{2\nu(t,r)}dt^{2}+e^{2\lambda(t,r)}\left[dr^{2}+r^{2}\left(d\theta^{2}
+\sin^{2}\theta d\phi^{2}\right)\right]\label{eqna01}
\end{equation}
where $\nu$ and $\lambda$ are the gravitational potentials.  We are investigating the
general case of a self-gravitating fluid in the presence of the electromagnetic field without
 placing arbitrary restrictions on the potentials.  For this model the Einstein equations
 are supplemented with Maxwell equations.  The Einstein field equations for a charged
 perfect fluid can be written as the system
\begin{subequations}
\label{eqn2}
\begin{eqnarray}
\rho&=&3\frac{\lambda_{t}^{2}}{e^{2\nu}}-\frac{1}{e^{2\lambda}}\left(2\lambda_{rr}
+\lambda_{r}^{2}+\frac{4\lambda_{r}}{r}\right)-\frac{E^{2}}{r^{4}e^{4\lambda}} \label{eqn01a}\\
p&=&\frac{1}{e^{2\nu}}\left(-3\lambda_{t}^{2}-2\lambda_{tt}+2\nu_{t}\lambda_{t}\right)
+\frac{1}{e^{2\lambda}}\left(\lambda_{r}^{2}+2\nu_{r}\lambda_{r}+\frac{2\nu_{r}}{r}
+\frac{2\lambda_{r}}{r}\right)+\frac{E^{2}}{r^{4}e^{4\lambda}} \label{eqn01b}\\
p&=&\frac{1}{e^{2\nu}}\left(-3\lambda_{t}^{2}-2\lambda_{tt}+2\nu_{t}\lambda_{t}\right)
+\frac{1}{e^{2\lambda}}\left(\nu_{rr}+\nu_{r}^{2}+\frac{\nu_{r}}{r}+\frac{\lambda_{r}}{r}
+\lambda_{rr}\right)-\frac{E^{2}}{r^{4}e^{4\lambda}} \label{eqn01c}\\
0&=&\nu_{r}\lambda_{t}-\lambda_{tr} \label{eqn01d}
\end{eqnarray}
\end{subequations}
Maxwell's equations yield
\begin{eqnarray}
E&=&r^{2}e^{\lambda-\nu} \Phi_{r}, \qquad E_{r} =\sigma r^{2}e^{3\lambda}\label{eqn02}
\end{eqnarray}
In the above $\rho$ is the energy density and $p$ is the isotropic pressure
 which are measured relative to the four-velocity $u^{a}=(e^{-\nu},0,0,0)$. Subscripts refer to partial derivatives with respect to that variable.
 The quantity $E=E(r)$ is an arbitrary constant of integration and $\sigma$
 is the proper charge density of the fluid.  We interpret $E$ as the total charge
 contained within the sphere of radius $r$ centred around the origin of the coordinate
 system.  Note that $\Phi_{r}=F_{10}$ is the only nonzero component of the electromagnetic
 field tensor $F_{ab}=\phi_{b;a}-\phi_{a;b}$ where $\phi_{a}=\left(\Phi(t,r),0,0,0\right)$.
 The Einstein-Maxwell system (\ref{eqn2})-(\ref{eqn02}) is a coupled system of equations in
 the variables $\rho$, $p$, $E$, $\sigma$, $\nu$ and $\lambda$.

The system of partial differential equations (\ref{eqn2}) can be simplified to produce an
underlying nonlinear second order equation.  Equation (\ref{eqn01d}) can be written as
\[
\nu_{r}=\left(\ln\lambda_{t}\right)_{r}
\]
Then (\ref{eqn01b}) and (\ref{eqn01c}) imply
\[
\left[e^{\lambda}\left(\lambda_{rr}-\lambda_{r}^{2}-\frac{\lambda_{r}}{r}\right)
+\frac{2E^{2}e^{-\lambda}}{r^{4}}\right]_{t}=0 \]
and the potential $\nu$ has been eliminated.  The Einstein field equations (\ref{eqn2})
can therefore be written in the equivalent form
\begin{subequations}
\label{eqn09}
\begin{eqnarray}
\rho&=&3e^{2h}-e^{-2\lambda}\left(2\lambda_{rr}+\lambda_{r}^{2}+\frac{4\lambda_{r}}{r}\right)
-\frac{E^{2}}{r^{4}e^{4\lambda}}  \label{eqn09a}\\
p&=&\frac{1}{\lambda_{t}e^{3\lambda}}\left[e^{\lambda}\left(\lambda_{r}^{2}
+\frac{2\lambda_{r}}{r}\right)-e^{3\lambda+2h}-\frac{E^{2}}{r^{4}e^{\lambda}}\right]_{t} \label{eqn09b}\\
e^{\nu}&=&\lambda_{t}e^{-h} \label{eqn09c}\\
e^{\lambda}\left(\lambda_{rr}-\lambda_{r}^{2}-\frac{\lambda_{r}}{r}\right)&=&
-\tilde{F}-\frac{2E^{2}}{r^{4}e^{\lambda}} \label{eqn09d}
\end{eqnarray}
\end{subequations}
for a  charged relativistic fluid.  In the above $h=h(t)$ and $\tilde{F}=\tilde{F}(r)$
are arbitrary constants of integration.  Equation (\ref{eqn09d}) is the
condition of pressure isotropy generalised to include the electric field.
 To find an exact solution of the field equations, we need to specify the
 functions $h$, $\tilde{F}$ and $E$ and solve equation (\ref{eqn09d})
 for $\lambda$.  We can then compute the quantities $\rho$ and $p$ from
 (\ref{eqn09a}) and (\ref{eqn09b}), and $\sigma$ follows from (\ref{eqn02}).

It is possible to write (\ref{eqn09d}) in a simpler form by eliminating
the exponential factor $e^{\lambda}$.  We use the transformation, first
introduced by Faulkes \cite{Faulkes} for neutral fluids, which has the adapted form
\begin{eqnarray*}
x&=&r^{2} \\
y&=&e^{-\lambda}\\
f(x)&=&\frac{\tilde{F}}{4r^{2}}\\
g(x)&=&\frac{E^{2}}{2r^{6}}
\end{eqnarray*}
Then (\ref{eqn09d}) becomes
\begin{equation}
y''=f(x)y^{2}+g(x)y^{3} \label{eqn013}
\end{equation}
which is the fundamental equation governing the behaviour of a shear-free charged fluid.  In (\ref{eqn013}) the primes denote the derivatives 
with respect to $x$.
Observe that (\ref{eqn013}) is a nonlinear partial differential equation since $y=y(t,x)$.
When $g=0$ then $y''=f(x)y^{2}$ for a neutral fluid which has been
studied by Maharaj \textit{et al} \cite{Maharaj} amongst others.

\section{Lie analysis}

We can verify that
\begin{equation}
G=a(x)\frac{\partial}{\partial x}+(b(x)y+c(x))\frac{\partial}{\partial y}\label{eqn12}
\end{equation}
is a symmetry of (\ref{eqn013}) and the relationship among the functions $a(x), b(x), c(x), f(x)$ and $g(x)$ is given by the following system of 
ordinary differential equations
\begin{subequations}
\begin{eqnarray}
a''&=&2b' \label{eqn13}\\
b''&=&2fc \label{eqn14}\\
c''&=&0 \label{eqn15}\\
a f'+(2a'+b)f&=&-3cg \label{eqn16}\\
a g'+(2a'+2b)g&=&0 \label{eqn17}
\end{eqnarray}
\end{subequations}
We now combine equations (\ref{eqn13})-(\ref{eqn17}) into one fifth order ordinary differential equation.  From (\ref{eqn13}) we have
\begin{equation}
2b=a'+\alpha \label{eqn18}
\end{equation}
where $\alpha$ is an arbitrary constant, from (\ref{eqn14}) we have
\begin{equation}
f=\frac{a'''}{4c} \label{eqn19}
\end{equation}
and from (\ref{eqn15}) we have
\begin{equation}
c=C_{0}+C_{1}x \label{eqn110}
\end{equation}
Previous analyses (Maharaj {\it et al} \cite{Maharaj} and Kweyama {\it et al} \cite{KMG}) have then solved  (\ref{eqn16}) and (\ref{eqn17}) for 
$f$ and $g$, respectively. Taking  (\ref{eqn19}) into account, Kweyama \textit{et al} \cite{KMG} obtained the fourth order 
integro-differential equation for $a$:
\begin{equation}
caa^{(iv)}+\left[c\left(\frac{5a'}{2}+\frac{\alpha}{2}\right)-c'a\right]a'''=-12g_{2}c^{3}a^{-3}\exp\left(-\int\frac{\alpha dx}{a}\right) \label
{oldeqna41}
\end{equation}
The nature of this particular integro-differential
equation has made it difficult to deduce much information in general.  Here, we focus on obtaining a fifth order 
purely {\it differential} equation.

We substitute (\ref{eqn18}) and (\ref{eqn19}) into (\ref{eqn16}) to obtain
\begin{equation}
g=-\frac{aa^{(iv)}}{12c^{2}}+\frac{aa'''c'}{12c^{3}}-\frac{5a'a'''}{24c^{2}}-\frac{\alpha a'''}{24c^{2}}\label{eqnn112}
\end{equation}
We then use (\ref{eqn18}) and (\ref{eqnn112}) in (\ref{eqn17}) to obtain the fifth order differential equation
\begin{eqnarray}
& &-\alpha^{2}c^{2}a'''-8\alpha c^{2}a'a'''-15c^{2}a'^{2}a'''+4\alpha a c c'a'''+18aca'c'a'''-6a^{2}c'^{2}a''' \nonumber\\
& &-5ac^{2}a''a'''-3\alpha ac^{2}a^{(iv)}-13ac^{2}a'a^{(iv)}+6a^{2}cc'a^{(iv)}-2a^{2}c^{2}a^{(v)}=0 \label{eqn112}
\end{eqnarray}
where $c(x)$ is given by (\ref{eqn110}). This is the first time that the symmetry analysis of (\ref{eqn013}) has reduced to solving a fifth order 
ordinary differential equation.  Given that the equation is purely differential, one can make recourse to numerical techniques if analytic 
solutions are elusive.

The fifth order equation (\ref{eqn112}) can be transformed into autonomous form via
the transformation
\begin{eqnarray}
X&=&\frac{C_{1}}{C_{0}+C_{1}x}, \qquad A= \frac{aC_{1}^{2}}{(C_{0}+C_{1}x)^{2}} \label{eqn113aa}
\end{eqnarray}
We end up with the following equation:
\begin{eqnarray}
& &2A^{2}A^{(v)}+13AA'A^{(iv)}-3\alpha AA^{(iv)}+5AA''A'''+15A'^{2}A'''\nonumber\\
& &-8\alpha A'A'''+\alpha^{2}A'''=0 \label{eqn123}
\end{eqnarray}
Once we obtain solutions to this equation, we can find $f$ and $g$ through direct substitution into (\ref{eqn19}) and (\ref{eqnn112}) 
respectively, after inverting (\ref{eqn113aa}).

We thus have the result that the equation
 \[y'' = \frac{a'''}{4c} y^2 +\left(-\frac{aa^{(iv)}}{12c^{2}}+\frac{aa'''c'}{12c^{3}}-\frac{5a'a'''}{24c^{2}}-\frac{\alpha a'''}{24c^{2}}\right) y^3 \]
admits the Lie point symmetry
\[G=a(x)\frac{\partial}{\partial x}+((a'(x)+\alpha)y/2+c(x))\frac{\partial}{\partial y}\]
where
\[c(x) = C_0+C_1 x\]
and $a(x)$ satisfies the fifth order ordinary differential equation
\begin{eqnarray*}
& &2A^{2}A^{(v)}+13AA'A^{(iv)}-3\alpha AA^{(iv)}+5AA''A'''+15A'^{2}A'''\nonumber\\
& &-8\alpha A'A'''+\alpha^{2}A'''=0 
\end{eqnarray*}
with
\begin{eqnarray*}
X&=&\frac{C_{1}}{C_{0}+C_{1}x}, \qquad A= \frac{aC_{1}^{2}}{(C_{0}+C_{1}x)^{2}} \label{eqn113a}
\end{eqnarray*}
As observed previously \cite{KMG}, this is a general result incorporating the {\it ad hoc} approach of Kweyama {\it et al} \cite{Kweyama} and 
Wafo Soh and Mahomed \cite{Wafo1}.  However, note that the fifth order equation (\ref{eqn123}) has not been derived previously.

If we let $\alpha=0$ in (\ref{eqn123}) we obtain
\begin{equation}
2A^{2}A^{(v)}+13AA'A^{(iv)}+5AA''A'''+15A'^{2}A'''=0 \label{eqn126}
\end{equation}
In carrying out the Lie analysis of (\ref{eqn126}), using PROGRAM LIE, we find that it has the following three symmetries
\begin{eqnarray*}
V_{1}&=&\frac{\partial}{\partial X} \label{eqn127a}\\
V_{2}&=&X\frac{\partial}{\partial X} \label{eqn127b}\\
V_{3}&=&A\frac{\partial}{\partial A} \label{eqn127c}
\end{eqnarray*}
We use $V_{1}$ to reduce the order of (\ref{eqn126}).  The variables for reduction are
\begin{eqnarray*}
u&=&A, \qquad v=A'
\end{eqnarray*}
and the reduced equation is
\begin{eqnarray}
& &2u^{2}v^{4}v^{(iv)}+14u^{2}v^{3}v'v'''+13uv^{4}v'''+8u^{2}v^{3}v''^{2}+22u^{2}v^{2}v'^{2}v''+57uv^{3}v'v''\nonumber\\
& &+15v^{4}v''+2u^{2}vv'^{4}+18uv^{2}v'^{3}+15v^{3}v'^{2}=0\label{eqn1210}
\end{eqnarray}
We analyse (\ref{eqn1210}) for symmetries, and obtain the following three symmetries using PROGRAM LIE:
\begin{eqnarray*}
X_{1}&=&u\frac{\partial}{\partial u}\\ \label{eqn1210a}
X_{2}&=&v\frac{\partial}{\partial v}\\ \label{eqn1210b}
X_{3}&=&2u^{2}\frac{\partial}{\partial u}+uv\frac{\partial}{\partial v} \label{eqn1210c}
\end{eqnarray*}
The symmetry $X_{3}$ determines the variables for reduction of (\ref{eqn1210})
\begin{eqnarray*}
r&=&u^{-(1/2)}v, \qquad s=u^{3/2}v'-\frac{1}{2}u^{1/2}v
\end{eqnarray*}
and the reduced equation in terms of new variables $r$ and $s$ is
\begin{equation}
2r^{4}s^{3}s'''+8r^{4}s^{2}s's''+14r^{3}s^{3}s''+2r^{4}ss'^{3}+22r^{3}s^{2}s'^{2}+22r^{2}s^{3}s'+2rs^{4}=0\label{eqn1211}
\end{equation}
The Lie symmetry analysis of (\ref{eqn1211}) gives three symmetries, namely
\begin{eqnarray*}
U_{1}&=&r\frac{\partial}{\partial r}\\ \label{eqn1211a}
U_{2}&=&s\frac{\partial}{\partial s}\\ \label{eqn1211b}
U_{3}&=&\frac{\partial}{\partial r}-\frac{s}{r}\frac{\partial}{\partial s} \label{1211c}
\end{eqnarray*}
The reduction variables generated by $U_{3}$ are
\begin{eqnarray*}
p&=&rs, \qquad q=rs'+s
\end{eqnarray*}
and the reduced equation is
\[2p^{3}q^{2}q''+2p^{3}qq'^{2}+8p^{2}q^{2}q'+2pq^{3}=0\]
which simplifies to
\begin{equation}
p^{2}qq''+p^{2}q'^{2}+4pqq'+q^{2}=0\label{eqn1213}
\end{equation}
Note that equation (\ref{eqn1213}) is in fact
\begin{equation}
\left(p^{2}q^{2}\right)''=0\label{eqn1214}
\end{equation}
and the solution of (\ref{eqn1214}) is
\begin{equation}
q^{2}=\frac{C_{2}}{p}+\frac{C_{3}}{p^{2}}\label{eqn1215}
\end{equation}
Note that this result contains the result obtained by Kweyama \textit{et al} \cite{KMG}.  Usually, when an $n$th order differential equation 
admits an $m$--dimensional Lie algebra of symmetries, with $m<n$, there is little hope for the solution of the equation via those symmetries.  
However, in this case we have been able to reduce the equation to quadratures due to the existence of hidden symmetries \cite{BSG, 
Abraham}.

We now return to (\ref{eqn112}) in the event that $\alpha\neq0$.  Integrating the equation (\ref{eqn112}) once yields the fourth order integro-
differential equation
\begin{equation}
caa^{(iv)}+\left[c\left(\frac{5a'}{2}+\frac{\alpha}{2}\right)-c'a\right]a'''=-12g_{2}c^{3}a^{-3}\exp\left(-\int\frac{\alpha dx}{a}\right) \label
{oldeqna41a}
\end{equation}
Integrating (\ref{oldeqna41a}) yields
\begin{eqnarray}
\frac{1}{2}a'''&=&2ca^{-(5/2)}\exp\left(-\int\frac{\alpha dx}{2a}\right)\left[f_{2}-3g_{2}\int ca^{-(3/2)}\right.\times\nonumber\\
& &\left.\exp\left(-\int\frac{\alpha dx}{2a}\right)dx\right]\label{oldeqna527}
\end{eqnarray}
where $f_{2}$ and $g_{2}$ are arbitrary constants of integration.  We multiply (\ref{oldeqna527}) by $a$ and then integrate to get
\begin{eqnarray}
\frac{1}{2}\int aa'''dx&=&2f_{2}I-6g_{2}\int \left[ca^{-(3/2)}\exp\left(-\int \frac{\alpha dx}{2a}\right)\times\right.\nonumber\\
& &\left.\int ca^{-(3/2)}\exp\left(-\int \frac{\alpha dx}{2a}\right)dx\right]dx+M\label{eqna528a}
\end{eqnarray}
where $M$ is an arbitrary constant of integration.  From (\ref{eqna528a}) we have
\begin{equation}
M=\frac{1}{2}aa''-\frac{1}{4}a'^{2}-2f_{2}I+3g_{2}I^{2} \label{eqna54}
\end{equation}
where \[I=\int ca^{-(3/2)}\exp \left(-\int \frac{\alpha dx}{2a}\right)dx\]
Again we multiply (\ref{oldeqna527}) by $aI$ and then integrate to obtain
\begin{eqnarray}
\frac{1}{2}\int aa'''Idx&=&2f_{2}\int ca^{-(3/2)}\exp\left(-\int\frac{\alpha dx}{2a}\right)Idx\nonumber\\
& &-6g_{2}\int ca^{-(3/2)}\exp\left(-\int\frac{\alpha dx}{2a}\right)I^{2}dx-N\label{eqntn}
\end{eqnarray}
where $N$ is an arbitrary constant of integration.  After performing the integrals in (\ref{eqntn}) we find that
\begin{eqnarray}
N&=&-a^{-(1/2)}\exp\left(-\int\frac{\alpha dx}{2a}\right)\left(ac'-\frac{1}{2}a'c+\frac{1}{2}\alpha c\right)\nonumber\\
& &-\left(\frac{1}{2}aa''-\frac{1}{4}a'^{2}+\frac{\alpha^{2}}{4}\right)I+f_{2}I^{2}-2g_{2}I^{3} \label{eqna55}
\end{eqnarray}
From (\ref{eqna54}) we have
\begin{equation}
\frac{1}{2}aa''-\frac{1}{4}a'^{2}=M+2f_{2}I-3g_{2}I^{2} \label{eqna541a}
\end{equation}
When substituting (\ref{eqna541a}) in (\ref{eqna55}) we obtain the following equation
\begin{eqnarray}
N&=&-a^{-(1/2)}\exp \left(-\int \frac{\alpha dx}{2a}\right)\left(ac'-\frac{1}{2}a'c+\frac{1}{2}\alpha\ c\right)-\left(M+\frac{\alpha^{2}}{4}\right)I
\nonumber\\
& &-f_{2}I^{2}+g_{2}I^{3}\label{eqna541b}
\end{eqnarray}

We thus have the result that the equation
\[ y'' = \frac{a'''}{4c} y^2 +\left( -\frac{aa^{(iv)}}{12c^{2}}+\frac{aa'''c'}{12c^{3}}-\frac{5a'a'''}{24c^{2}}-\frac{\alpha a'''}{24c^{2}} \right)y^3 \]
admits the Lie point symmetry
\be
G=a(x)\frac{\partial}{\partial x}+((a'(x)+\alpha)y/2+c(x))\frac{\partial}{\partial y} \label{sym}
\ee
where
\[ c(x) = C_0+C_1 x \]
and $a(x)$ is constrained by the first order integro--differential equation
\[N=-a^{-(1/2)}\exp \left(-\int \frac{\alpha dx}{2a}\right)\left(ac'-\frac{1}{2}a'c+\frac{1}{2}\alpha c\right)-\left(M+\frac{\alpha^{2}}{4}\right)I-f_{2}I^
{2}+g_{2}I^{3}\]
and $M$, $N$, $f_{2}$ and $g_{2}$ are all arbitrary constants of integration.  This is the first time that the admittance of symmetry by (\ref
{eqn013}) has been reduced to solving essentially a first order equation.
This is a general result incorporating both the {\it ad hoc} approach of Kweyama {\it et al} \cite{Kweyama} and the Noether symmetry results 
of Wafo Soh and Mahomed \cite{Wafo1}.

In the event that $\alpha=0$, we have that (\ref{eqna541b}) reduces to
\be
N=-a^{-(1/2)}\left(ac'-\frac{1}{2}a'c\right)-MI-f_{2}I^{2}+g_{2}I^{3} \label{a0eqn}
\ee
this time with
\[I=\int ca^{-(3/2)}dx\]
We can now integrate (\ref{a0eqn}) and obtain the same result as in Wafo Soh and Mahomed \cite{Wafo1}. See Kweyama {\it et al} \cite
{KMG} for a full discussion of this case.

It remains to solve (\ref{eqna541b}) when $\alpha\neq0$.
If we set
\begin{eqnarray}
X&=&\frac{C_{1}}{C_{0}+C_{1}x} \qquad A= \frac{aC_{1}^{2}}{(C_{0}+C_{1}x)^{2}} \label{eqn113}
\end{eqnarray}
then (\ref{eqna541b}) reduces to
\begin{equation}N=C_{1}\left[-A^{-(1/2)}\exp\left(\int\frac{\alpha dX}{2A}\right)\left(\frac{A'}{2}+\frac{\alpha}{2}\right)\right]+C_{1}\left(M+\frac
{\alpha^{2}}{4}\right)I-C_{1}^{2}f_{2}I^{2}-C^{3}_{1}g_{2}I^{3} \label{for}
\end{equation}
where, in this case,
\[I=\int A^{-(3/2)}\exp\left(\int\frac{\alpha dX}{2A}\right)dX\]
We can now integrate (\ref{for}) further to obtain
\begin{equation}
A^{1/2}\exp\left(\int\frac{\alpha dX}{2A}\right)=\int\left[-\frac{N}{C_{1}}+\left(M+\frac{\alpha^{2}}{4}\right)I-C_{1}f_{2}I^{2}-C^{2}_{1}g_{2}I^
{3}\right]dX+P \label{asoln}
\end{equation}
which is an implicit solution for (\ref{eqn112}) with $\alpha\neq0$.  We note that no reductions of (\ref{eqna541b}) (or its fourth order 
counterpart) have been previously found for nonzero $\alpha$.

Having found $a$ via (\ref{asoln}) and (\ref{eqn113}) we now focus on (\ref{eqn013}).  Using (\ref{sym}) we can transform (\ref{eqn013}) into 
the autonomous form
\begin{equation}
Y''+\alpha Y'+\left(M+\frac{\alpha^{2}}{4}\right)Y=f_{2}Y^{2}+g_{2}Y^{3}+N \label{eqna53}
\end{equation}
via the transformation
\begin{eqnarray*}
X&=&\int\frac{dx}{a}, \qquad Y=y \exp\left(-\int\frac{b dx}{a}\right)-\int\frac{c}{a}\exp\left(-\int\frac{b dx}{a}\right)dx
\end{eqnarray*}
The form (\ref{eqna53}) can be more directly analysed.  However,
when $\alpha\neq 0$, we find that we cannot directly reduce (\ref{eqna53}) to quadratures.  Further (Lie) analysis yeilds that, when $f_
{2}\neq0$, $g_{2}\neq0$ then (\ref{eqna53})
 has the following two symmetries
\begin{eqnarray}
G_{1}&=&\frac{\partial}{\partial X} \label{eqna59a}\\
G_{2}&=&e^{(\alpha/3)X}\frac{\partial}{\partial X}-e^{(\alpha/3)X}\left(\frac{\alpha}{3}Y+\frac{\alpha f_{2}}{9g_{2}}\right)\frac{\partial}{\partial 
Y} \label{eqna59b}
\end{eqnarray}
provided the following conditions are satisfied
\begin{equation}
M=-\frac{f_{2}^{2}}{3g_{2}}-\frac{\alpha^{2}}{36}, \qquad N=\frac{f_{2}^{3}}{27g_{2}^{2}}-\frac{2\alpha^{2}f_{2}}{27g_{2}} \label{mnconds}
\end{equation}
Here, $G_2$ is a new symmetry -- $G_1$ is just the transformed form of (\ref{sym}).
If we further transform (\ref{eqna53})  using $G_2$ we obtain
\[{\cal Y}''=g_{2}{\cal Y}^{3} \]
with  solution
\[{\cal X}-{\cal X}_{0}=\int \frac{d{\cal Y}}{\sqrt{\frac{g_{2}}{2}{\cal Y}^{4}+{\cal C}}} \]
where
\begin{eqnarray*}
\cal X&=&-\frac{3}{\alpha}e^{-(\alpha/3)X},  \qquad {\cal Y}=e^{(\alpha/3)X}\left(Y+\frac{f_{2}}{3g_{2}}\right)
\end{eqnarray*}
As noted previously \cite{KMG},  the values in (\ref{mnconds}) correspond directly to a simplification of the eigenvalue problem associated 
with a dynamical systems analysis of (\ref{eqna53}).

\section{Discussion}

We have reduced the system of Einstein-Maxwell field equations to a single nonlinear second order partial differential equation modeling the 
behaviour of spherically symmetric charged fluids.  This was achieved by utilising the generalised transformation due to Faulkes \cite
{Faulkes}.  The master equation is a partial differential equation because $y=y(x,t)$.  However it may be treated as an ordinary differential equation.  We then 
derived (for the first time) a fifth order nonlinear differential equation.  The solutions of this fifth order equation yield the solution to the 
governing Einstein-Maxwell field equations.  The equation was then transformed into an autonomous form.  In the case of $\alpha\neq0$, we 
performed a Lie symmetry analysis of the autonomous fifth order differential equation.  This allowed us to find a solution to  the equation.  
On letting $C_{2} =2$ in (\ref{eqn1215}), we regained the result obtained by Kweyama \textit{et al} \cite{KMG}. In our approach we have 
eliminated the need to solve a nonlinear integro-differential equation which was a necessary feature of earlier analyses.

For the $\alpha\neq0$ case, we were not able to make progress via a Lie symmetry analysis of the autonomous fifth order equation.  
However we were able to integrate the non-autonomous fifth order differential equation directly.  We obtained an implicit solution for this 
case - a solution that has not been obtained previously.

~\\
{\bf Acknowledgements}\\
MCK and KSG thank the National Research Foundation and the
University of KwaZulu-Natal for financial support. SDM acknowledges
that this work is based upon research supported by the South African
Research Chair Initiative of the Department of Science and
Technology and the National Research Foundation.
~\\


\newpage
\begin{thebibliography}{9}
\bibitem{Ivanov}
B. V. Ivanov, Phys. Rev. D \textbf{65}, 104001 (2002).

\bibitem{Sharma}
R. Sharma, S. Mukherjee and S. D. Maharaj, Gen. Relativ. Gravit.
 \textbf{33}, 999 (2001).

\bibitem{Mak}
M. K. Mak and T. Harko,  Int. J. Mod. Phys. 
D  \textbf{13}, 149 (2004).

\bibitem{Komathiraj}
K. Komathiraj and S. D.  Maharaj,  
Int. J. Mod. Phys. D
\textbf{16}, 1803 (2007).

\bibitem{Lasky}
P. D. Lasky and A. W. C. Lun,  Phys. Rev. D \textbf{75}, 024031 (2007).

\bibitem{Lun}
P. D. Lasky and A. W. C. Lun, Phys. Rev. D
\textbf{75}, 104010 (2007).

\bibitem{Di}
A. Di Prisco, L. Herrera, G. Le Denmat, M. A. H. MacCallum and N. O. Santos,
Phys. Rev. D \textbf{76}, 064017 (2007).

\bibitem{Herrera}
L. Herrera, A. Di Prisco, E. Fuenmayor and O. Troconis,  
Int. J. Mod. Phys. D
\textbf{18}, 129 (2009).

\bibitem{Kustaanheimo}
P. Kustaanheimo and B. Qvist, 
Soc. Sci. Fennica, Comment. Phys. - Math.
\textbf{XIII}, 12 (1948).

\bibitem{Srivastava}
D. C. Srivastava, 
Class. Quantum Grav. \textbf{4}, 1093 (1987).

\bibitem{Suss89}
R. A. Sussman,  Gen. Relativ.
Gravit.  \textbf{21}, 1281 (1989).

\bibitem{Krasinski}
A. Krasinski, Inhomogeneous cosmological models (Cambridge University Press, Cambridge, 1997).

\bibitem{Suss88a}
R. A. Sussman,  J.  Math. Phys. 
\textbf{29}, 945 (1988).

\bibitem{Suss88b}
R. A. Sussman, 
J. Math. Phys. 
\textbf{29}, 1177 (1988).

\bibitem{SrivaDC}
D. C. Srivastava, 
Fortschr. Phys.  \textbf{40}, 31 (1992).

\bibitem{Faulkes}
M. Faulkes,  Can. J.  Phys.  \textbf{47}, 1989  (1969).

\bibitem{KMG}
M. C. Kweyama, K. S. Govinder and S. D. Maharaj, 
Class. Quantum Grav. \textbf{28}, 105005 (2011).

\bibitem{Maharaj}
S. D. Maharaj, P. G. L. Leach and R. Maartens,   Gen. Relativ.
Gravit.  \textbf{28}, 35 (1996).

\bibitem{Kweyama}
M. C. Kweyama, S. D.  Maharaj and K. S. Govinder,  Nonlinear Analysis: RWA
submitted (2011).

\bibitem{Wafo1}
C. Wafo Soh and F. M. Mahomed,  
Class. Quantum Grav. \textbf{17}, 3063 (2000).

\bibitem{BSG}
B. Abraham--Shrauner and A. Guo, J. Phys. A: Math. Gen.  {\bf 25}, 5597 (1992).

\bibitem{Abraham}
B. Abraham-Shrauner,  J. Math.
Phys. \textbf{34}, 4809 (1993).

\end{thebibliography}
\end{document}